\documentclass[aps,apl,twocolumn,showpacs,amsmath,floatfix,superscriptaddress]{revtex4}

\usepackage{graphicx}

\begin{document}

\title{Enhanced Many-Electron Effects on Excited States of Gated Bilayer Graphene}

\author{Yufeng Liang, Li Yang}
\affiliation{Department of Physics, Washington University in St.
Louis, St. Louis, MO 63136, USA}

\date{\today}

\begin{abstract}
By employing the first-principles GW-Bethe-Salpeter Equation (BSE)
simulation, we obtain, for the first time, the accurate
quasiparticle (QP) band gap, optical absorption spectra and their
dependence on the gate field of gated bilayer graphene (GBLG).
Many-electron effects are shown to be extremely important to
understand these excited-state properties; enhanced
electron-electron interactions dramatically enlarge the QP band
gap; infrared optical absorption spectra are dictated by bright
bound excitons. Our results well explain recent experiments and
satisfyingly solve the puzzle about the inconsistency between
experimentally measured transport and optical band gaps. Moreover,
our calculation reveals fine excitonic structures and predicts
exotic excitonic effects that have not been observed yet, which
can be of interest for optoelectronics applications based on GBLG.
\end{abstract}


\maketitle

Despite its intriguing electronic, thermal and optical properties
\cite{2004MonoG, 2007Graphene, review-1}, intrinsic graphene is a
gapless semimetal, impeding its utility in bipolar devices and
subsequent broad applications. Therefore, huge efforts have been
made to overcome this barrier by generating a finite band gap in
graphene or its derivatives \cite{2007BBG1, 2007BBG2, Graphane,
Fluorographene}. One promising approach is to apply the gate
electric field perpendicular to the AB (Bernal) stacked bilayer
graphene (BLG) to break the inversion symmetry of sublattices
\cite{2007BBG1, 2007BBG2, 2007Transport1, 2009Optical1,
2009Optical2}. Such an induced band gap of GBLG can be tuned in a
wide range by field strength \cite{2009Optical1, 2007Transport1,
2010Transport2, 2012Transport3}, offering an important degree of
freedom to optimize performance of graphene devices.

However, we are still in lack of a satisfying understanding on
fundamental properties of GBLG, such as its quasiparticle (QP)
band gap and optical excitations. For instance, electrical
conductance experiments \cite{2007Transport1, 2010Transport2,
2012Transport3} have confirmed the existence of a finite QP band
gap but their measured value is annoyed by the inevitable contact
resistance between electrodes and graphene sheet; while
noncontacting optical measurements \cite{2009Optical1,
2009Optical2, 2009kim} have revealed a tunable band gap in GBLG
but these results are indirect because the optical absorption peak
(edge) is not conceptually equivalent to the QP band gap
\cite{BSE, review-BSE}. Particularly, enhanced excitonic effects
often dramatically shift the optical absorption peak as we have
seen in many other reduced dimensional semiconductors
\cite{Catalin2004, Feng2005}, making this inconsistence even more
serious. Therefore, an accurate calculation with many-electron
effects is urgent to settle the above inconsistency.

Conventional density functional theory (DFT) simulations cannot
answer above questions because of their known deficiencies of
handling excited-state properties \cite{BSE, review-BSE, GW}.
Tight-binding models \cite{TBGWBSE} have revealed appealing
physics of excitons in GBLG, but it has to rely on parameters. In
particular, recent \emph{ab initio} GW-BSE simulation has
successfully predicted enhanced many-electron effects on intrinsic
graphene \cite{BSEGraphene}, which are confirmed by subsequent
experiments \cite{graphene-1, graphene-2, graphene-3}. Therefore,
a reliable first-principles calculation with many-electron effects
included is also promising.

More importantly, beyond providing reliable parameters for device
design, obtaining knowledge about excited states of GBLG will be
of fundamental interest because it will fill our knowledge gap on
many-electron interactions in two-dimension (2-D) narrow-gap
semiconductors, a field that has not been well understood yet. In
fact, it is challenging for first-principles simulations to
accurately capture the nearly metallic electronic screening of
narrow-gap semiconductors. For this purpose, an improved algorithm
has to be developed and shall be of broad interest for the
electronic-structure community.

In this Letter, with the modified model accurately describing the
screening, we conclude three important remarks about excited
states of GBLG. 1) The QP band gap and its dependence on the gate
field are obtained. The self-energy correction is significant
because of the enhanced electron-electron (\emph{e-e})
interactions; the calculated QP band gaps are more than 150 \% of
previous DFT predictions \cite{LDA1, LDA2}, which is beneficial
for device applications since a wider band gap implies higher
working temperature. 2) Optical absorption spectra of GBLG are
dominated by excitonic effects. With electron-hole (\emph{e-h})
interactions included, our calculated absorption peaks are in
excellent agreement with recent experiments \cite{2009Optical1},
well explaining the inconsistency between QP gap and optical gap.
3) Excitons in GBLG exhibit a number of unusual features. For
example, the electron and hole of the lowest-energy dark exciton
are completely separated onto two graphene layers, giving rise to
an optical approach to polarize BLG. Moreover, this separation of
electron and hole offers a neat opportunity to evaluate entangling
effects, such as the exchange interaction, of many-electron
systems.

To reveal the significance of many-body correlations in GBLG, we
perform the calculation of excited states as the following
procedure. First, ground-state energy and wavefunctions are
obtained by DFT within the local density approximation (LDA).
Secondly, the QP energy is calculated within the single-shot $GW$
approximation \cite{GW}. Finally we obtain the exciton energy,
wavefunctions and optical absorption spectra by solving the
following BSE \cite{BSE}
\begin{equation}
(E_{c\mathbf{k}}-E_{v\mathbf{k}})A^S_{vc\mathbf{k}}+\sum_{v^{\prime}c^{\prime}\mathbf{k}^{\prime}}\\
\langle vc\mathbf{k}\vert K^{eh} \vert
v^{\prime}c^{\prime}\mathbf{k}^{\prime}\rangle
A^S_{v^{\prime}c^{\prime}\mathbf{k}^{\prime}}=\Omega^S
A^S_{vc\mathbf{k}} \label{BSE-1}
\end{equation}
where $A^S_{vc\mathbf{k}}$ is the amplitude of excitonic state $S$
consisting of single-particle hole state $\vert
v\mathbf{k}\rangle$ and electron state $\vert c\mathbf{k}\rangle$.
$K^{eh}$ is the \emph{e-h} interaction kernel and $\Omega^S$ is
the exciton excitation energy. $E_{c\mathbf{k}}$ and
$E_{v\mathbf{k}}$ are QP energy of electrons and holes,
respectively.

All calculations are based on plane-wave basis and norm-conserving
pseudopotential approximations with a 60-Ry energy cutoff. To
eliminate the spurious interaction between neighboring BLG, the
slab-truncation scheme is applied to mimic isolated GBLG
\cite{Truncation1, Truncation2}. The electric field is applied by
the periodic sawtooth potential perpendicular to graphene layers.

The crucial part of describing many-electron interactions is to
obtain the dielectric function. For GBLG with the truncated
Coulomb interaction, the inverse static dielectric function
$\epsilon^{-1}(\mathbf{q})$ rapidly changes within the long
wave-length regime $\mathbf{q}\rightarrow 0$, which is similar to
what have been noticed in recent first-principles simulations of
carbon nanotubes (CNTs) \cite{Catalin2004}. A brute-force way to
capture this feature is to use a dense q-grid, which demands
formidable computational resources. To solve this problem, we
deliberately employ the mini Brillouin zone (BZ) sampling scheme
to count in this sharply-varying character as motivated by Refs.
\cite{Truncation1, Truncation2}, and extend it to both evaluating
the QP energies and solving the BSE (see the supplementary
material for details). As a result, a $72\times72\times1$ coarse
k-grid sampling is adequate for the GW calculation. In addition,
we employ a partial $1440\times1440\times1$ fine k-grid sampling
around the Dirac cone for a dependable BSE calculation.

\begin{figure}
\centering
\includegraphics*[scale=0.75]{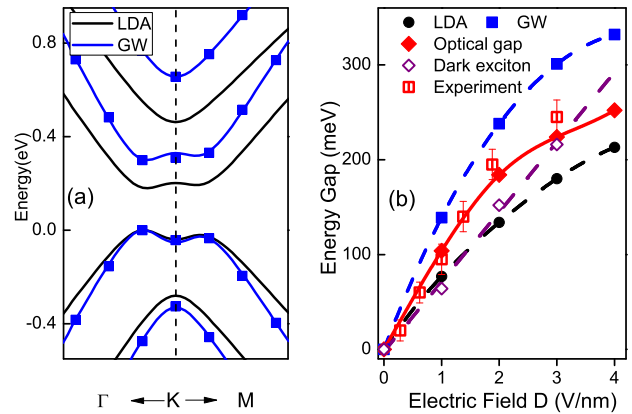}
\caption{(Color online) (a) DFT/LDA and GW calculated band
structures around the Dirac point of BLG under a gate field of 2
V/nm. (b) Comparison of the "gap" values obtained from different
approaches and their dependence on the gate electric field. The
value of the optical gap is defined by the position of the first
bright peak of the optical absorption spectrum. The experimental
values are extracted from Ref.\cite{2009Optical1}} \label{bandgap}
\end{figure}

The LDA and $GW$ band structures near the BZ corner (the K point)
are plotted in Fig. \ref{bandgap} (a) for GBLG, respectively. The
applied gate field induces a finite band gap and changes the band
dispersion to the Mexican-hat feature. After including the
self-energy correction via the GW calculation, the
Mexican-hat-shaped feature remains intact, nevertheless the
fundamental band gap is significantly enlarged due to the
depressed screening of isolated GBLG. Moreover, the slope of band
dispersion is sharpen by the self-energy correction, implying a
smaller effective mass of free carriers.

We also investigate the QP band gap dependence on the applied gate
field as shown in Fig. \ref{bandgap} (b). The QP band gap can be
varied from zero up to 300 meV under experimentally reachable gate
field, which is also more than 150$\%$ of previous DFT
predictions. This larger tunable range and a wider gap are desired
for device applications because a wider gap means higher working
temperature and lower noise. Moreover, if listing the ratio of the
self-energy correction to their DFT/LDA value, we see it is 56\%,
67\%, 78\% and 81\%, respectively, as the applied field is
decreased. This growing trend of the correction ratio for the
smaller gap is of particular interest because recent experiments
\cite{2012Transport3} shows a possible small band gap (around a
few meV) even for BLG in the absence of gate field. However, due
to the limit of the accuracy of our numerical simulation, we
cannot resolve those energy differences below 10 meV and hence
more advanced simulation technique needs to be developed.

In addition, the recent optical measurements of the optical gap
are plotted in Fig. \ref{bandgap} (b) as well. The key feature is
that the QP gap is substantially larger than both previous DFT
predictions \cite{LDA1, LDA2} and measurements from optical
experiments, \cite{SCFTB, 2009Optical1}. The inconsistency QP band
gap with optical measurements has also been observed in several
other semiconducting nanostructures \cite{Catalin2004, Feng2005},
which manifests enhanced excitonic effects and motivates the
following calculation on optical spectra of GBLG.

\begin{figure}
\includegraphics*[scale=1.1]{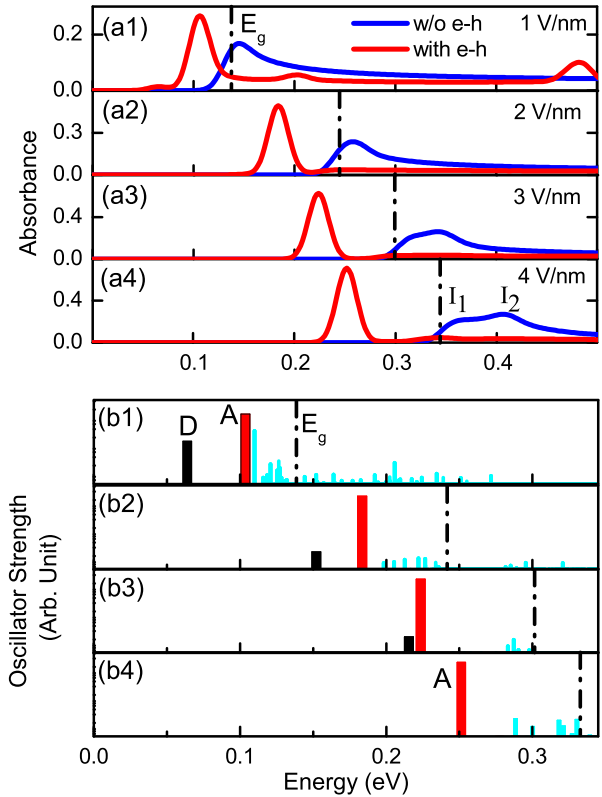}
\caption{(Color online) (a) Optical absorption spectra of GBLG.
The vertical black dashed line indicates the GW fundamental gap.
The incident light is polarized parallel to the graphene plane. A
10 meV Gaussian smearing is applied. (b) Optical activity and
eigenenergy of excitons. Each bar represents one exciton state and
its height (plotted in the logarithmic scale) indicates the
corresponding optical activity. The lowest-energy dark exciton $D$
and the prominent exciton $A$ are particularly outlined by widened
dark and red bars, respectively.} \label{spectra}
\end{figure}

Fig. \ref{spectra} depicts the optical absorption spectrum and its
evolution subject to the increasing field magnitude. We first
focus on absorption spectra in the absence of the \emph{e-h}
interactions (blue lines in Figs. \ref{spectra} (a)). In the
low-energy regime, the absorption is mostly contributed by the
transition from the highest valence band (v1) and the lowest
conduction band (c1). As expected, the absorption onset displays a
blueshift as the electric field uplifts the band gap magnitude.
Meanwhile, the prominent absorption feature is gradually broadened
and split into a double-peak structure ($I_1$ and $I_2$) which
stems from the two one-dimensional-like von-Hove singularities
\cite{TBGWBSE, MultiLayer} at opposite \lq\lq Mexican-hat
brims\rq\rq (Fig. \ref{bandgap} (a)), which is consistent with
previous DFT results \cite{LDA1}.

\begin{figure}
\includegraphics*[scale=0.35]{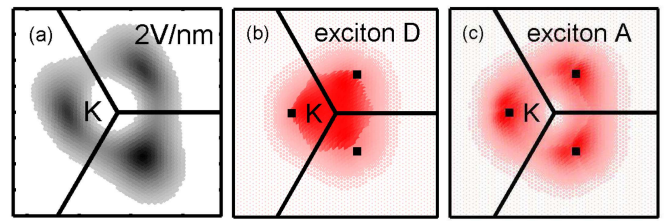}
\caption{(Color online) (a) The distribution of single-particle
oscillator strength in the reciprocal space. We only include
transitions from the highest valence band to the lowest conduction
band. (b) and (c) The distributions of the square of the exciton
amplitude ($\vert A^S_{vc\mathbf{k}}\vert^2$) of the dark exciton
$D$ and bright exciton $A$ in the reciprocal space. The square
black dots mark the three identical locations of the minimum
energy gap.} \label{DIST}
\end{figure}

Surprisingly, the von-Hove singularity at the K point, however,
does not contribute greatly to the absorption and therefore is not
resolved in the spectra. This is because the relevant valence
state $\vert v\mathbf{k}\rangle$  and the conduction state $\vert
c\mathbf{k}\rangle$ at the Dirac point K are strongly localized on
different layers upon field-induced symmetry breaking. Therefore,
the overlap of wavefunctions is very small, which eventually leads
to a negligible oscillator strength. This can be seen in Fig.
\ref{DIST} (a), in which we present the contour plot of the
oscillator strength around the corner of the first BZ. The
strongest oscillator strength is actually from the \lq\lq
Mexican-hat brims\rq\rq regime while it is almost zero at the K
point.

With \emph{e-h} interactions included, a different optical
absorption picture emerges. As shown in Figs. \ref{spectra} (a),
the exciton effect dramatically reshapes the spectra; the broad
asymmetric absorption peak in the single-particle picture is
replaced by a symmetric prominent absorption peak. This peak lies
below the QP band gap, indicating the existence of bound
\emph{e-h} pairs. \emph{The binding energy under four sampling
voltages is 35, 54, 76 and 80 meV,} respectively, which is also
close to previous tight-binding calculations \cite{TBGWBSE}.
Remarkably these peak-position is in excellent agreement with the
previous infrared microspectroscopy experiment \cite{2009Optical1}
as shown in Fig. \ref{bandgap} (b). Under realistic experimental
setups, both self-energy corrections and e-h interactions shall be
reduced by the screening effect of dielectric substrates. On the
other hand, these reductions may cancel each other more or less
\cite{2011yang}, resulting in such a good match of our
calculations with experimental data.

In the higher energy regime (around 0.4 eV) next to the first
optical active peak, the absorbance maintains a constant on the
whole ($\sim$ 3$\%$), which is significantly smaller than 4.6\%,
the ideal value of the optical absorbance of BLG \cite{optical-2,
BGConstant2}. This is due to the sum rule of oscillator strength
\cite{BSE} because \emph{e-h} interactions drains the absorbance
from the high-energy regime to enhance the exciton peak.

It has to be pointed out that electron-phonon coupling shall be
another important factor in determining the infrared optical
spectra of GBLG. For example, a G-mode phonon at 195meV has been
found to be in Fano interference coupled with e-h excitations in
GBLG \cite{2010wang}. Therefore, we may expect this dip feature
from such a G-mode phonon may impact the lineshape of our studied
exciton absorption peaks.

A close inspection of solutions of the BSE reveals an intriguing
exciton picture that have not been observed by experiments. We
plot the oscillator strength of excitons in a logarithmic scale in
Figs. \ref{spectra} (b). The isolated exciton state with the
largest oscillator strength, $A$, is responsible for the symmetric
prominent absorption peaks in the spectra. Surprisingly, there is
one lower-lying excion, $D$, with a much weaker oscillator
strength for most gated fields (except 4V/nm). This is contrary to
the usual effective-mass model, in which the lowest singlet
exciton shall be the brightest one involved with two bands.

Furthermore, we observe that both the position and oscillator
strength of this dark exciton $D$ are more sensitive to the gate
field than those of bright exciton $A$. As plotted in Figs.
\ref{spectra} (b), the energy of $D$ progressively approaches that
of $A$ with an increasing gate field strength and its optical
activity is strongly quenched simultaneously. In particular, when
the gate field is more than 3V/nm, the order of the bright and
dark excitons is switched as shown in Fig. \ref{bandgap} (b). This
tunable energy difference can surely affect the thermal population
of exciton states and their luminescent performance. Such a
tunable energy order of excitons is in qualitative agreement with
previous tight-binding studies \cite{TBGWBSE}.

To understand the brightness of these exciton states, we need to
further investigate the origin of their optical activity. For a
typical field strength (2 V/nm), Figs. \ref{DIST} (b) and (c)
display the distribution of the square of exciton amplitude
$A_{vc\mathbf{k}}^S$ for those two interested excitons $A$ and
$D$. Since the optical activity of an exciton $i$  \cite{BSE} is
\begin{equation}
|\langle 0 | \vec{v}| i \rangle|^2=|\sum_{vck}A^i_{vck}\langle vk
| \vec{v} | ck \rangle|^2, \label{absorption}
\end{equation}
which roughly means the product of the single-particle oscillator
strength shown in Fig. \ref{DIST} (a) and exciton amplitude shown
in Figs. \ref{DIST} (b) or (c), we immediately see the product of
exciton $D$ is much bigger than exciton $A$, suggesting their
markedly different brightness.

\begin{figure}
\includegraphics*[scale=0.55]{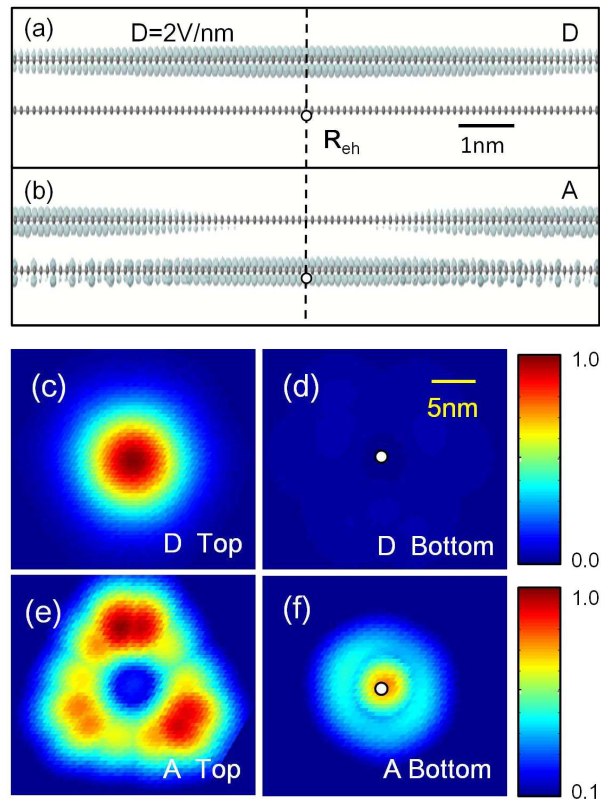}
\caption{(Color online) (a) and (b) Side views of the isosurface
plot of the square of wavefunctions of the excitons $D$ and $A$.
(c) to (f) Top view of these exciton wavefunctions for top and
bottom layers, respectively. The hole position is marked by the
open circle in (a) and (b) while it is located at the center of
the bottom layer in (d) and (f).} \label{EXCITON}
\end{figure}

Fig. \ref{EXCITON} visualizes exciton wavefunctions in the real
space. As is readily seen, both excitons $A$ and $D$ are strong
charge transfer excitons but with distinct characters. In
particular, the electron and hole of the dark exciton $D$ almost
become disentangled. As shown in Figs. \ref{EXCITON} (a), (c) and
(d), the electron and hole wavefunctions of exciton $D$ are nearly
completely separated into two layer. This is very dissimilar to
the \emph{e-h} correlation in other 2-D semiconductors
\cite{GraphaneXcton, BSEGraphene}. From the perspective of
optoelectronic application, exciton $D$ could yield interesting
possibility of efficient \emph{e-h} separation and polarize BLG by
optical excitations. For the exciton $A$, the degree of \emph{e-h}
separation is much lower. In Figs. \ref{EXCITON} (b), (e) and (f),
the electron distributes over a ring on the top layer while on the
bottom layer the electron distributes on a disk centered at the
hole.

Moreover, these excitonic wavefunctions will be crucial to
understand why the dark exciton $D$ and the dark exciton $A$
respond very differently to the electric field. As concluded in
Fig. \ref{bandgap} (b), the energy level of exciton $D$ exhibits
an approximately linear relationship with the field strength,
whereas that of exciton $A$ shows a nonlinear behavior. This can
be rationalized by the fact that exciton $D$ can be viewed as a
plane of dipole composed of dissociated electron and hole as
revealed in Fig .\ref{EXCITON} (a), whose energy levels of the
positive and negative poles linearly depend on the applied gate
field. In contrast, the electron and hole for the exciton $A$ are
spatially entangled and therefore the energy level is less
sensitive to the gate field and does not follow a simple linear
trend. This explains the origin of the energy order switch when
the applied gate field is more than 3 V/nm.

In conclusion, we have provided first-principles calculations for
the QP energy and excitonic effects of GBLG. \emph{E-e} and
\emph{e-h} interactions are significant and must be considered to
understand the electronic structure and optical excitations of
GBLG. Moreover, our calculation well explains recent experiments
and reveals more physics associated with many-electron effects.
Finally, we have observed an exotic dark exciton structure that is
not likely to present in conventional direct band gap
semiconductors. The different degree of charge transfer for
different exciton states may be useful in optoelectronic
applications.

The computational resources have been provided by Lonestar of
Teragrid at the Texas Advanced Computing Center. The ground state
calculation is performed by the Quantum Espresso. The GW-BSE
calculation is done with the BerkeleyGW package \cite{BGW} with
minor modifications for excited states calculations in GBLG.

\end{document}